\documentclass[twocolumn,showpacs,preprintnumbers,amsmath,amssymb]{revtex4}

\usepackage{graphicx}% Include figure files
\usepackage{dcolumn}% Align table columns on decimal point
\usepackage{bm}% bold math
\usepackage{color}
\usepackage{amsmath}

\begin{document}

\title{Multi-threshold second-order phase transition}% Force line breaks with \\

\author{Wei Zhuang$^{1}$, Deshui Yu$^{1} $, Zhiwen Liu$^{2}$, Jingbiao Chen$^{1}$}\email{jbchen@pku.edu.cn}

\address{$^{1}$Institute of Quantum Electronics, and State Key
Laboratory of Advanced Optical Communication System $\&$ Network,
School of Electronics Engineering $\&$ Computer Science, Peking
University, Beijing 100871, P. R. China\\
$^{2}$Department of Electrical Engineering,The Pennsylvania State
University, University Park, PA 16802, USA}
\date{\today}

\begin{abstract}
We present a theory of the multi-threshold second-order phase
transition, and experimentally demonstrate the multi-threshold
second-order phase transition phenomenon. With carefully selected
parameters, in an external cavity diode laser system, we observe
second-order phase transition with multiple (three or four)
thresholds in the measured power-current-temperature three
dimensional phase diagram. Such controlled death and revival of
second-order phase transition sheds new insight into the nature of
ubiquitous second-order phase transition. Our theory and experiment
show that the single threshold second-order phase transition is only
a special case of the more general multi-threshold second-order
phase transition, which is an even richer phenomenon.
\end{abstract}


\maketitle

Landau's theory~\cite{Landau1,Landau2} of the second order phase
transition (SOPT) is crucial for a variety of important physical
systems, including helium
superfluidity~\cite{Landau3,Landau4,Landau5},
superconductivity~\cite{Ginzburg}, ferromagnetic
system~\cite{Onsager,Kaufman}, Bose-Einstein
condensates~\cite{Pitaevskii}, as well as
lasers~\cite{Frederick,Haken,DeGiorgio,Corti,Haken2,Haken3, Scully}.
However, all of these SOPTs observed in experiments or investigated
in theories so far possess only a single critical point (or
threshold). In other words, there is only one critical point for
symmetry breaking~\cite{Landau1}.

In this Letter, we present a theory of the multi-threshold SOPT, and
demonstrate experimentally  the multi-threshold SOPT phenomenon in
an external cavity diode laser system with carefully selected
parameters. Our results show that the single threshold SOPT is only
a special case of the more general multi-threshold SOPT. Our theory
and experiment imply that previous studied fascinating SOPTs, such
as spontaneous magnetization~\cite{Ising},
superfluidity~\cite{Kapitza,Tisza} and
superconductivity~\cite{Onnes,Cooper,Bardeen}, which are solely
limited to single critical-point behavior, may be extended to the
multi-threshold SOPT.

The Landau expression for the free energy $F$ near the critical
point~\cite{Landau1,Landau2}, a function of an order parameter
$\psi$,
\begin{equation}\label{FreeEnergy1}
F=F_{n}+\alpha|\psi|^{2}+\frac{\beta}{2}|\psi|^{4}
\end{equation}
is generally applied to all aforementioned SOPTs. Here, the
coefficient $\alpha=\alpha_{c}(T-T_{c})$, is assumed by Landau to be
a simple linear function of temperature T when T is lower than the
critical temperature $T_{c}$. $F_{n}$ is the free energy of the
normal phase.

However, if the coefficient $\alpha$ consists of periodic function
of T, for example, a trigonometric function, the system can evolve
to a multi-threshold SOPT phenomenon with suitable experimental
parameters, as demonstrated in this Letter. In this case, the free
energy is expressed as,

\begin{equation}\label{FreeEnergy2}
F=F_{n}+\alpha_{c}(T-T_{c}-T_{mc}cos\phi(T))|\psi|^{2}+\frac{\beta}{2}|\psi|^{4}
\end{equation}

\begin{figure}
%\Requires \usepackage{graphicx}
\includegraphics[width=9cm]{fig1a.eps}\\
\includegraphics[width=8cm]{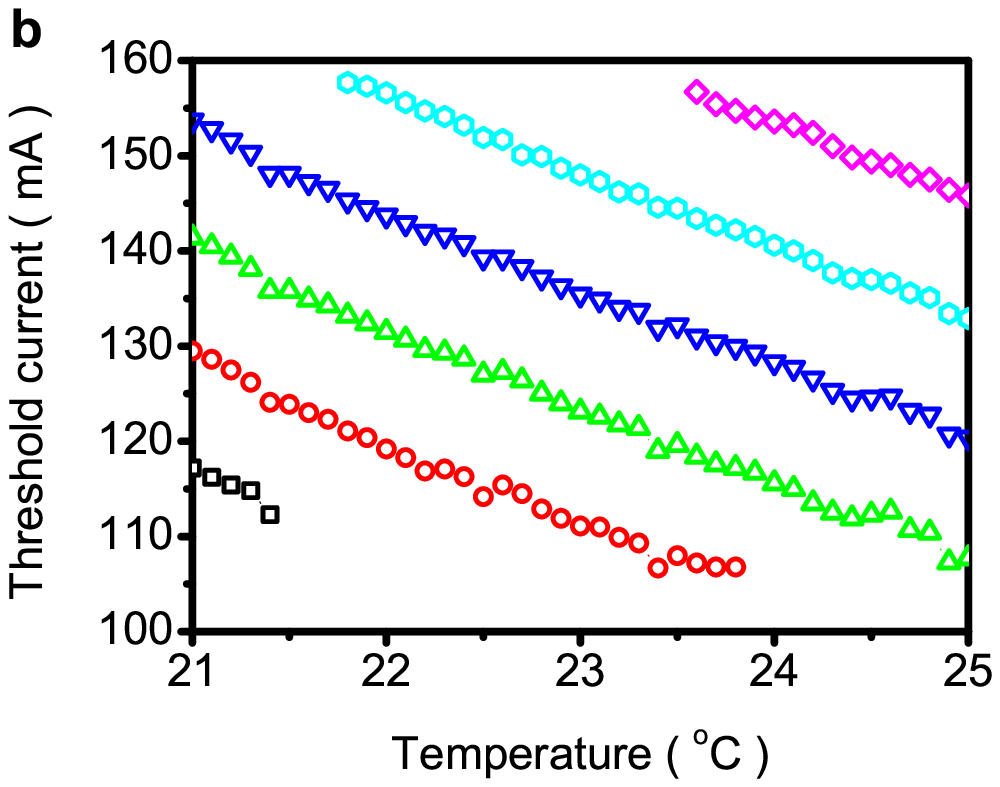}\\
\includegraphics[width=9cm]{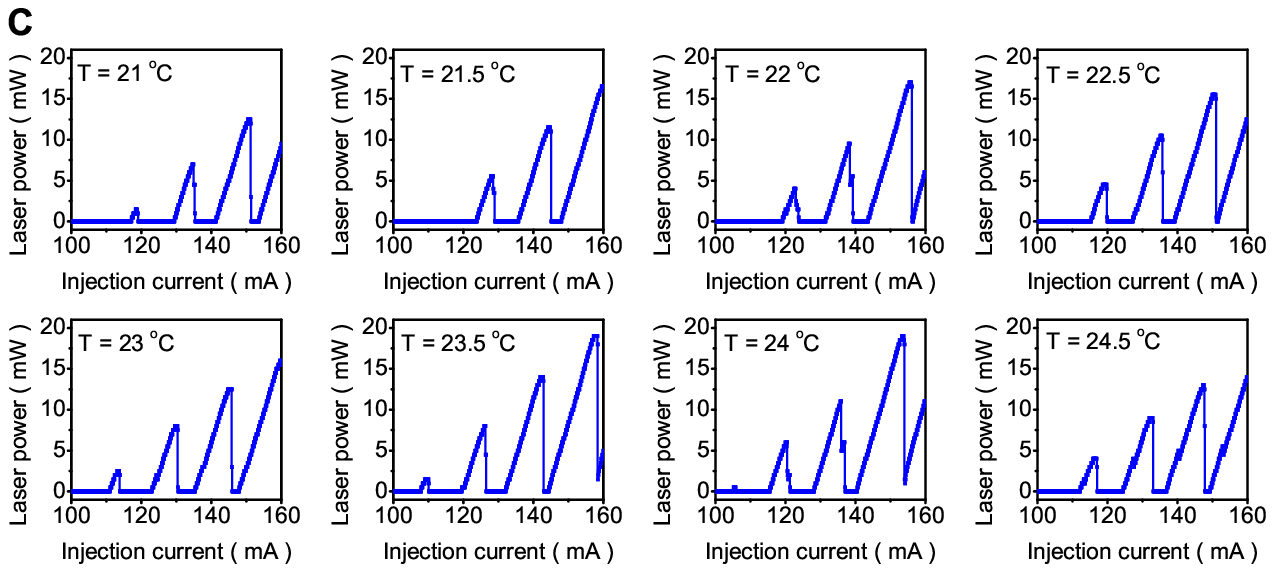}\\
\caption{(a) The measured three-dimensional  phase diagrams of
multi-threshold laser power as a function of current and diode
temperature. (b) Relation curves between threshold current and
temperature of ECDL. Different data symbols represent relevant
thresholds current of the multiple lasing regions showed in (a). A
linear relationship can be observed, and the vertical displacements
between adjacent lines are almost equal to 12.3mA. (c) Detailed
power-current curves at different diode temperatures. Injection
current is tuned with a step of 0.1mA from 100mA to 160mA limited.
There are more than one injection current thresholds which decrease
while the temperature of ECDL is increased.}\label{Figure1}
\end{figure}

Each period of the cosine function of T produces a threshold. As we
will see later, in this case the order parameter $\psi$ will be a
function with a periodic pattern of T. If  $\phi(T)$ only weakly
depends on the temperature T, Eq.(2) is reduced to Eq.(1).

The lasing action of a laser can be considered as a Bose-Einstein
condensation of photons, which also belongs to the SOPT. The
analogies with superconductivity and ferromagnet, developed by
Haken, Scully and others
~\cite{Frederick,Haken,DeGiorgio,Corti,Haken2,Haken3,Scully}, show
that the effective Ginzburg-Landau free energy of laser is,
\begin{equation}\label{FreeEnergy3}
 G=G_{0}-\frac{1}{4}(A-C)|u|^{2}+\frac{1}{8}B|u|^{4}\\
=G_{0}-\frac{1}{4}(A-C)n+\frac{1}{8}B|n|^{2}
\end{equation}
Where the amplitude of electric field $u$ corresponds to the order
parameter $\psi$ in Eq.(1) and Eq.(2), $G_{0}$ corresponds to
$F_{n}$. A is the gain in the active medium, B is the saturation
coefficient, C is the cavity loss rate, and the photon density in
laser cavity is $n$.

In order to experimentally demonstrate the multi-threshold SOPT in a
laser, the key challenge is to carefully set the gain and cavity
loss term A-C in Eq.(3) to be a periodic function of the effective
temperature (the population inversion in laser corresponds to
negative temperature~\cite{Ramsey}), which is related to injection
current I in a diode laser.

Since the refractive index of semiconductor chip depends on the
injection current(see Eq.(12)), it can be shown that for a diode
laser system,
\begin{equation}\label{PhotonNumber1}
A-C=\alpha_{c}(I-I_{c}-I_{{mc}}cos\phi(I))
\end{equation}

In Landau's theory, the dependence of the order parameter on the
effective temperature near the critical point is determined by the
condition that the free energy as a function of the order parameter
is minimized ~\cite{Landau1,Landau2,Ginzburg}. For a laser near the
threshold, from Eq.(3) we obtain
\begin{equation}\label{PhotonNumber2}
n=\frac{A-C}{B}
\end{equation}

Near and above threshold, the exact solution of the photon density
obtained from general laser rate equations involving carrier density
N and photon density n in a diode laser,
\begin{equation}\label{RateEquation1}
\frac{dN}{dt}=\frac{I}{qV}-\frac{N}{\tau_{N}}-g_{N}(N-N_{tr})n
\end{equation}
\begin{equation}\label{RateEquation2}
\frac{dn}{dt}=\Gamma g_{N}(N-N_{tr})n-\frac{n}{\tau_{p}}
\end{equation}
should be,
\begin{equation}\label{PhotonNumber3}
n=\frac{A}{C}\frac{A-C}{B}
\end{equation}
Here the gain A and saturation coefficient B are
\begin{equation}\label{Photon2}
A=\frac{\Gamma g_{N}\tau_{N}}{qV}I-\Gamma g_{N}N_{tr}
\end{equation}
\begin{equation}\label{Photon3}
B=\frac{\Gamma (g_{N}\tau_{N})^{2}}{qV}I-\Gamma
(g_{N})^{2}\tau_{N}N_{tr}
\end{equation}
Particularly, the cavity loss rate C is a trigonometric function of
I and is given by
\begin{equation}\label{Photon4}
C=C_{1}-C_{2}cos\phi(I)
\end{equation}
\begin{equation}\label{Photon7}
\phi(I)=\phi_{0}+\frac{\alpha_{H} l_{d}}{\nu_{g}\Gamma}
C+\frac{2\omega l_{d}}{c}\frac{\partial \eta}{\partial I} I
\end{equation}
where, $C_{1}=\nu_{g}\alpha_{i}+\frac{\nu_{g}}{l_{d}}\ln R_{c}$,
$C_{2}=\frac{\nu_{g}}{l_{d}}\frac{R_{m}}{R_{c}}$,
$R_{c}=\sqrt{\frac{1-R_{2}R_{3}}{R_{1}(R_{3}-R_{2})}}$,
$R_{m}=\sqrt{\frac{R_{2}}{R_{1}}}\frac{1-R_{3}}{R_{3}-R_{2}}$, q is
the electron charge, $\phi_{0}$ is a constant phase, and other
symbols are listed in Table 1. Equation (13) describes the effect of
injection current through thermal effect and threshold gain through
carrier density on the refractive index~\cite{Sargent, Coldren}.

It should be noted that, Eq.(8) has the same form as the exact
solution of laser quantum theory~\cite{Scully}. This expression of
Eq.(8) can be reduced to Eq.(5) when laser is near threshold,
i.e.~A$\approx$C. The measured output power of laser is hence given
by P=$\kappa C n h \nu V_{p}$, where $\kappa$ is the output coupling
efficiency, $h\nu$ is the photon energy, $V_{p}$ is the light field
mode volume.

In an external cavity diode laser (ECDL) system,with carefully
selected parameters for realizing a desired $\phi(I)$ in A-C
expression, we observed SOPT with multiple (three or four)
thresholds in the measured power-current ($P$-$I$) relationship,
which corresponds to $\psi$-T relation derived from Eq. (1) in
Landau's theory. The gradual evolvement from the SOPT with multiple
critical points to single critical point SOPT confirmed
experimentally and agrees well with Eq.(3) and Eq.(4).
Figure~\ref{Figure1}a shows the measured three-dimensional (3D)
phase diagrams of multi-threshold SOPT of the ECDL's output power as
a function of the diode temperature $T_{d}$ and the injection
current $I$. It can be clearly seen that up to four lasing
thresholds can exist at a certain operating diode temperature and
that the laser revives at a new threshold after each "death". The
multiple threshold currents as a function of the operating diode
temperature are shown in Fig.~\ref{Figure1}b. A linear dependence of
threshold current on diode temperature $T_{d}$ can be observed, and
the vertical displacement between two adjacent lines are almost 12.3
mA. The thresholds of the injection current $I$ decrease as the
diode temperature $T_{d}$ is increased, which is in contrast to that
in a usual ECDLs~\cite{Sargent, Coldren} where the threshold current
increases with the diode temperature. In Fig.~\ref{Figure1}c, we
show the detailed phase diagrams of our laser system (the measured
power-current curves at different diode temperatures).

\begin{table}
\caption{\label{tab:table 1}Typical parameters of semiconductor
diode~\cite{Sargent,Coldren} used in calculations.}
\begin{ruledtabular}
\begin{tabular}{lll}
  Parameter&Description&Value\\
  \hline
  $\Gamma$ & Optical confinement factor& 0.44\\
  $g_{N}$ & Gain coefficient & 3 $\times$ $10^{-12}$ $m^{3}$/$s$ \\
  $\tau_{N}$ & Carrier lifetime & 3 ns\\
  $l_{d}$ & Active region length & $3\times 10^{-4} m$ \\
  $V$ & Active region volume & $6 \times10^{-17} m^{3}$\\
  $N_{tr}$ & Transparency carrier density& $2 \times 10^{24} m^{-3}$ \\
  $\nu_{g}$& Group light speed & $7 \times 10^{7} m/s$\\
  $\alpha_{i}$ & Internal material loss & $1000 m^{-1}$\\
  $\eta_{0}$ & Refractive index& 3.5\\
  $\alpha_{H}$& Linewidth enhancement factor & 6  \\
  $R_{1}$& Facet reflectivity & 0.3 \\
  $R_{2}$ &AR coated facet reflectivity & $5 \times 10^{-3}$ \\
  $R_{3}$& External cavity reflectivity & 0.05 \\
  $\frac{\partial \eta}{\partial I}$ & Refractive index coefficient  & 0.1 \\
  $\lambda$&Output wavelength& 830 nm\\
\end{tabular}
\end{ruledtabular}
%\end{table*}
\end{table}

In our experiment, GaAs diode laser at a wave-length
$\lambda=830~nm$ with the output facet coated with a multi-layer
antireflection coating was operated with the standard
Litterman-Metcalf external cavity configuration. The size of the
active region of laser diode is
0.1$\mu$m$\times$2$\mu$m$\times$300$\mu$m
(height$\times$width$\times$length). The distance between laser
diode and optical grating over a range from 35 mm to 1000 mm does
not affect the results. The calculations shown in Fig.3b are based
the laser parameters listed in Table 1, and the nonlinear gain
effect has been included in calculation.

\begin{figure}
\includegraphics[width=8cm]{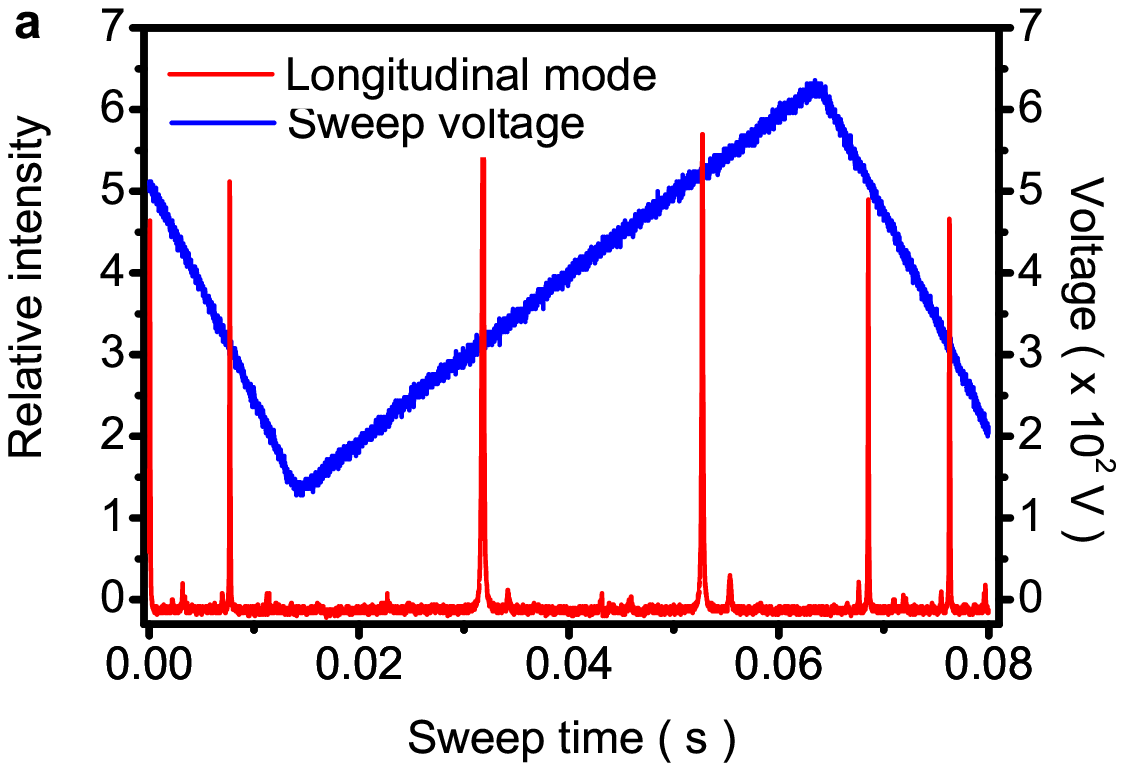}\\
\includegraphics[width=8cm]{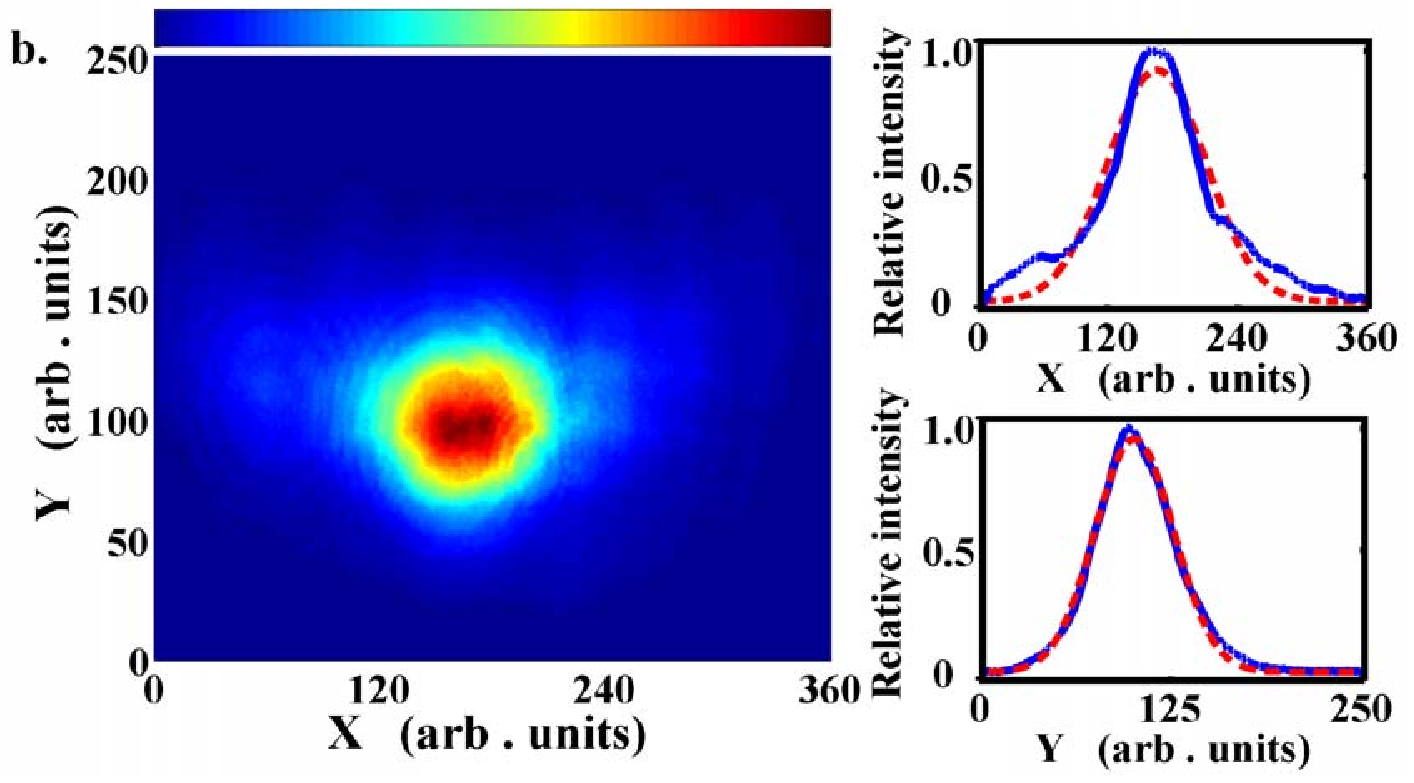}\\
\caption{(a) The longitudinal mode of ECDL measured by scanning a
high-fineness Fabry-Perot etalon. The blue line represents the
sweeping voltage, and the red line denotes the longitudinal mode,
which clearly shows that our ECDL is operating at single
longitudinal mode. (b) The transverse mode profile of ECDL measured
by a Laser Beam Profile Analyzer (LBP-2-USB) at half a meter away
from the laser. The beam profile is shown in left picture, while the
distributions of the relative intensity in x direction and y
direction are shown in right figures (blue lines) with the Gaussian
fitting curves (red lines). The measured waist of laser beam in x
direction is 879.93$\mu$m in the Gaussian fit, while the waist in y
direction is 478.61$\mu$m in the Gaussian fit.}\label{Figure2}
\end{figure}

In the same diode laser, measurements during the last 38 months show
exactly the same multi-threshold SOPT behavior. Once the laser is
"alive", it operates at a single longitudinal mode, and the laser
wavelength is identical at the beginnings of different thresholds.
The beam profile of our ECDL is shown in Fig.~\ref{Figure2}, which
clearly shows that it operates at both single longitudinal mode and
fundamental Gaussian traverse mode. When the laser is at "death"
state, the measured fluorescence spectrum is as wide as 20 nm. We
have also measured the $P$-$I$ curves at different lasing
wavelengths within the 20 nm gain range, and the multi-threshold
SOPT consistently exists.

\begin{figure}
\includegraphics[width=8cm]{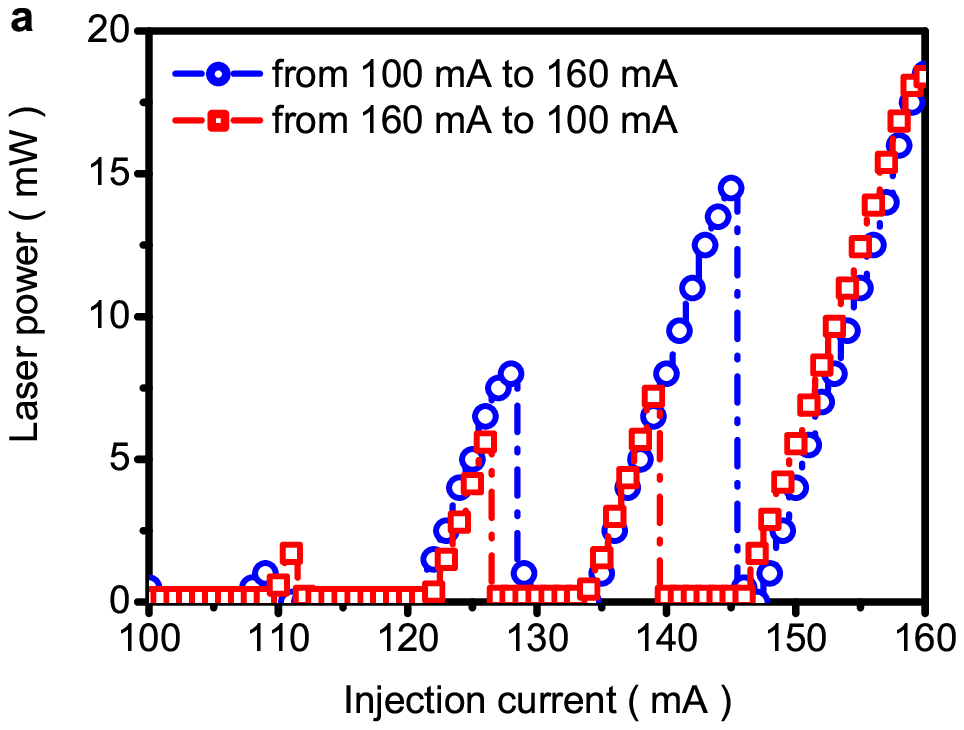}\\
\includegraphics[width=8cm]{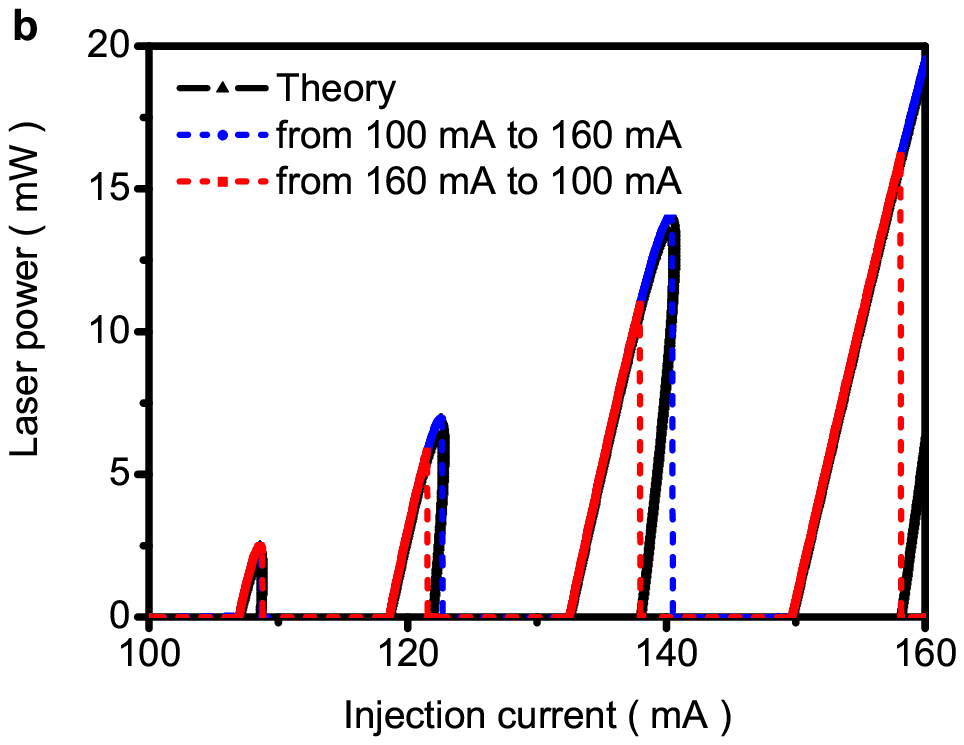}\\
\caption{(a) Experimentally measured power-current dependence at
diode temperature 23$^{o}$C. Blue circles were recorded when
injection current was increased from 100 mA to 160 mA while red
squares correspond to a decreasing injection current from 160 mA to
100 mA. Hysteresis can be clearly observed. (b) Calculated results;
blue and red lines are deduced from the calculated curve (black)
which simulates the experimental results in (a). Theory could well
explain the multi-threshold phenomenon and the optical bi-stability
effect (Hysteresis) by comparing (a) and (b).}\label{Figure3}
\end{figure}

Fig.~\ref{Figure3}a shows the dependence of the ECDL's output power
as a function of the injection current $I$ at diode temperature
$T_{d}=23^{o}C$. The corresponding theoretical result obtained by
soloving Eq.~(\ref{PhotonNumber3}) is given in Fig.~\ref{Figure3}b.
The blue curves of both figures are the measured laser power as the
injection current increased from 100~mA to 160~mA, while the red
curves are the measured results when the injection current is
decreased. Four thresholds exist in both cases, with the first
threshold appearing at $I$=107~mA (there is no threshold below
100~mA injection current). Each laser threshold is followed with a
bi-stable state, which corresponds to the first-order phase
transition~\cite{Scott}. In our case, when the injection current is
gradually increased, the laser begins to lase at each threshold, and
its power drops to zero almost suddenly after reaching a maximum
value. However, if the injection current is decreased, the lasing
action begins at a point that is lower than the previous maximum,
and the output power decreases linearly to zero at the same
threshold.

Although Landau's theory of the SOPT~\cite{Landau1,Landau2} and
Ginzburg-Landau equations of superconductivity~\cite{Ginzburg,
Rosenstein} have been used to describe laser
action~\cite{Haken,DeGiorgio,Corti,Haken2,Haken3, Scully,
Staliunas}, all fascinating SOPT explained so far in numerous
physical systems such as spontaneous magnetization~\cite{Ising},
superfluidity~\cite{Kapitza,Tisza} and
superconductivity~\cite{Onnes,Cooper,Bardeen}, are solely limited to
single critical point category. We believe that the SOPT with
multiple critical points described by Eq.(2) is more general and
even richer phenomenon than the conventional SOPT. The single
threshold behavior is only a special case of the multi-threshold
SOPT as Eq.(2) can be reduced to Eq.(1) if $\phi(T)$ does not depend
on the temperature T. The simple laser rate equations with the gain
and cavity loss term A-C depending linearly on population inversion
only describes single-threshold
SOPT~\cite{Haken,DeGiorgio,Corti,Haken2,Haken3, Scully, Christopher}
or multi-threshld of first-order phase transition~\cite{Scully,
Christopher}. We have also observed the multi-threshold SOPT
phenomenon in other diode lasers operating at a wavelength of 852
nm. The key is to carefully select the parameters to satisfy the
gain and loss condition shown in Eq.(4).

In summary, we proposed and demonstrated that once the effective
temperature periodically affects the order parameter, a physical
system described by Landau's theory of SOPT can be extended to
exhibit multi-threshold SOPT. Our findings lead to a natural
question: can multi-threshold SOPT also emerge in other physical
systems, such as superconductivity, superfluidity, ferromagnetic
system, Bose-Einstein condensates etc., which usually display single
SOPT as in a laser?

\textbf{Acknowledge:} The authors thank Yiqiu Wang, Zhichao Xu for
helpful discussions. This work is supported by the National Natural
Science Foundation of China under No. 10874009 and No.11074011.

\end{document}